\documentclass[numberedappendix,apj]{emulateapj}
\usepackage{apjfonts}
\usepackage{amssymb,amsmath}
\usepackage{natbib}
\usepackage{graphicx}
\shorttitle{Tsallis statistics}
\shortauthors{Esquivel \& Lazarian}
\slugcomment{Draft version, \today}

\begin{document}

\title{Tsallis statistics as a tool for studying interstellar turbulence}

\author{A. Esquivel\altaffilmark{1} and A. Lazarian\altaffilmark{2}}
\affil{
\altaffilmark{1}Instituto de Ciencias Nucleares, Universidad Nacional
  Aut\'{o}noma de M\'{e}xico, Apartado Postal 70-543, 04510 M\'{e}xico
  D.F., M\'{e}xico\\
\altaffilmark{2}Astronomy Department, University of Wisconsin-Madison, 475
    N.Charter St., Madison, WI 53706-1582, USA}

\email{esquivel@nucleares.unam.mx,lazarian@astro.wisc.edu}

\begin{abstract}

We used magnetohydrodynamic (MHD) simulations of interstellar
turbulence to study the probability distribution functions (PDFs) of
increments of density, velocity, and magnetic field. We found that the
PDFs are well described by a Tsallis distribution, following the same
general trends found in solar wind and Electron MHD studies. We found
that the PDFs of density are very different in subsonic and supersonic
turbulence. In order to extend this work to ISM observations we
studied maps of column density obtained from 3D MHD simulations. From
the column density maps we found 
the parameters that fit to Tsallis distributions and
demonstrated that these parameters vary with the sonic and Alfv\'en
Mach numbers of turbulence. This opens avenues for using Tsallis
distributions to study the dynamical and perhaps magnetic states of interstellar gas. 

\end{abstract}

\keywords{ISM: general --- MHD  --- turbulence}

\section{Introduction}
\label{sec:introduction}

Turbulence is known to be a crucial ingredient in many astrophysical
phenomena that take place in the interstellar medium (ISM): such as
star formation, cosmic ray dispersion, formation and evolution of the
(also very important) magnetic field, and virtually every transport
process (see for instance the reviews of \citealt{2004ARA&A..42..211E,
  2004RvMP...76..125M, 2007prpl.conf...63B, 2007ARA&A..45..565M}, and
references therein).
This has brought many researchers to the study of turbulence, from
both observational and theoretical perspectives.

From the observational point of view there are several techniques
aimed to study properties of ISM turbulence. Scintillation studies 
\citep[see for instance][]{1989MNRAS.238..963N,1990ApJ...353L..29S}
have been very successful to characterize ISM turbulence, however
they are limited to density fluctuations in only ionized media. 
For neutral media the density fluctuations can also be obtained from
column density maps \citep[see][]{2003ApJ...583..308P}.
Other avenue of research is provided by radio spectroscopic observations,
using for instance the centroids of spectral lines
\citep{1951ZA.....30...17V, 1958RvMP...30.1035M, 1985ApJ...295..479D,
 1985ApJ...295..466K, 1987ApJ...317..686O, 1994ApJ...436..728F,
 1994ApJ...429..645M, 1995ApJ...450L..27M, 1998ApJ...504..889L,
 1999ApJ...524..895M}, or linewidth maps 
\citep{1981MNRAS.194..809L,1992MNRAS.256..641L, 1984ApJ...277..556S,
  1987ASSL..134..349S}. Spectroscopic observations have the advantage
of containing information about the underlying turbulent velocity
field. However, they are also sensitive to fluctuations in density, and the
separation of the two contributions alone has proved to be a difficult
problem \citep[see][]{2006AIPC..874..301L}. 

Much of the effort to relate observations with models of turbulence
has focused on obtaining the spectral index (log-log slope of the
power-spectrum) of density and/or velocity.
To this purpose several methods have been developed and tested through
numerical simulations. Among them, we can mention the
``Velocity Channel Analysis''
\citep[VCA,][]{2000ApJ...537..720L, 2003MNRAS.342..325E,
  2004ApJ...616..943L, 2001ApJ...555..130L, 2003ApJ...583..308P,
  2009ApJ...693.1074C}, the   
``Spectral Correlation Function''
 \citep[SCF][]{1999ApJ...524..887R,
  2001ApJ...547..862P}\footnote{The VCA and SCF uses the same sort of
  two point statistics. An insubstantial difference is that VCA
  stresses the use of fluctuation spectra, while SCF uses structure
  functions. A more fundamental difference stem from the fact that the
  VCA is based on the analytical theory
  relating the observed spectrum of intensity fluctuations and the
  underlying spectra of velocity and density, while the choice of
  structure function normalization adopted in the SCF prevents the use
  of the developed theoretical machinery for a quantitative analysis
  \citep{2009SSRv..143..357L}.}, ``Velocity coordinate spectrum'' 
\citep{2008ApJ...686..350L, 2006ApJ...652.1348L}, or ``Modified
Velocity Centroids'' \citep{2003ApJ...592L..37L, 2005ApJ...631..320E,
  2006A&A...452..223O,2007MNRAS.381.1733E, 2006astro.ph.11462C,
  2009ApJ...693.1074C, 2009arXiv0910.1384P}
Different methods are suitable in different situations, for instance
VCA is adequate to retrieve velocity information from supersonic
turbulence, while centroids are only  useful in subsonic (or at most,
transonic) turbulence. 

More sophisticated techniques for the analysis of the data, including
using wavelets instead of Fourier transforms
\citep{1994Ap&SS.221...25H, 1998A&A...336..697S} have been
attempted. A rather sophisticated example of such analysis is a use of
the Principal Component Analysis\footnote{We should mention, however,
  that the use of integral transforms, other than the the Fourier
  transform, provide a new way of getting the same information. For
  instance, using Fourier transforms one can also attempt to get the
  empirical relation between the underlying velocity and the
  statistics of the fluctuations.
  However, the VCA-obtained relations  are very difficult 
  to match empirically because of the strong non-linear nature of the
  transforms. Therefore it is not surprising that the empirically
  obtained relations for the PCA do not reproduce important cases
  discovered by the VCA. The PCA analysis failed to reveal the
  dependence of the Position-Position-Velocity (PPV) fluctuations on
  the density, which should be the case for the shallow spectrum of
  fluctuations. See more on the testing the PPV dependencies in
  \citet{2009ApJ...693.1074C}.}(PCA)
\citep{1997ApJ...475..173H, 2002ApJ...566..276B}.

The techniques above provide the spectrum of turbulent velocity and/or
density. Spectra, however, do not provide a complete description of
turbulence. For  example, \citet{2008ApJ...688.1021C} showed that
fields with a very 
different distribution of density could have the same power spectrum.
Intermittency and topology are some of the turbulence features that
require additional statistical tools. 
Measures of turbulence intermittency include probability distribution
functions (PDFs, see for instance \citealt{1994ApJ...436..728F, 
  1996ApJ...463..623L, 1998ApJ...504..889L, 2000ApJ...535..869K}),
 higher-order She-L\'ev\^eque exponents
\citep{1994PhRvL..72..336S, 2003ApJ...583..308P, 2005AIPC..784..299F,
  2008A&A...481..367H}. 
In addition,  
{\it genus}  \citep[see also][]{2002ASPC..276..182L,
2007ApJ...663..244K, 2008ApJ...688.1021C} was used to study the
topological structure of the diffuse gas distribution.  

Astrophysical turbulence is very complex, e.g. it can happen in the
multi-phase media with numerous energy sources. Therefore   
the synergy of different techniques provide an invaluable insight into
the properties of the ISM turbulence and the self-organization of
interstellar medium. For instance, 
\citet{2007ApJ...658..423K} and \citet{2009ApJ...693..250B}
have applied several statistical measures, including a measure of
bispectrum, to a wide set of MHD simulations and
synthetic-observational data. More recent 
work applying the same set of techniques to the Small Magellanic Cloud
(SMC) data revealed the advantages of such an approach
\citep{2010BSLK}. It is the consistency of the results obtained by
different techniques, which made the study of turbulence properties in
SMC more reliable.  

Thus it is important to proceed with the quest for new techniques for
astrophysical studies (see \citealt{2009SSRv..143..357L} and
ref. therein).
In this paper we add the Tsallis statistics to the arsenal of tools
available for ISM studies of turbulence. This is a description of the
PDFs that have been successfully used previously in the context of solar wind
observations and it shares similarities to previous studies of
ISM turbulence \citep[e.g.][]{2000ApJ...535..869K}. 

The earlier solar wind relation between the Tsallis statistics and
turbulence was sought on the basis of 1D MHD simulations. We know that
the properties of turbulence strongly depend on its dimensions. Thus
we use 3D MHD simulations.
We analyze a reduced set of MHD simulations, including
column density maps which are the most easily available data from
observations.

This paper is organized as follows. In section \ref{sec:tsallis-intro}
we briefly present the Tsallis distribution, which we will use to
characterize magnetohydrodynamic (MHD) simulations. The description of
the numerical models and the results (for 3D data as well as column
density) can be found in sections
\ref{sec:tsallis-mhd}--\ref{sec:column}. We discuss the 
significance  of our findings in \ref{sec:discussion}. Finally, in
section \ref{sec:summary}  we provide our summary. 

\section{Tsallis Statistics}
\label{sec:tsallis-intro}

\citet*{2004GeoRL..3116807B,2004JGRA..10912107B, 2005PhyA..356..375B,
  2005JGRA..11007110B,  2006ApJ...642..584B, 2006ApJ...644L..83B,
  2006PhyA..361..173B, 2007JGRA..11207206B, 2007ApJ...668.1246B,
  2009ApJ...691L..82B},
 have studied the PDFs of
the variation of the magnetic field strength in the solar wind, measured
by the {\it Voyager 1} and {\it 2} spacecrafts. In particular,
they analyzed the fluctuations of magnetic field increments
$[B(t+\tau_m)-B(t)]$ as a function of the timescale $\tau_m$, and showed
that for a wide range of scales their PDFs could be described by a
Tsallis (or $q$-Gaussian) distribution
\citep{1988JSP....52..479T}. 
The Tsallis  PDF of increments ($\Delta f$) of an arbitrary function has the form:
\begin{equation}
  R_q=A \left[1+\left(q-1\right)\frac{\Delta f\left(t,\tau\right)^2}{w^2}
    \right]^{-1/(q-1)},
\label{eq:tsallis}
\end{equation}
where $q$ is the so called ``entropic index'' or ``non-extensivity
parameter'' and is related to the size of the tail of the
distribution, $w$ provides a measure of the with of the PDF (related
to the dispersion of the distribution), while
$A$ measures the amplitude.
Notice that the Tsallis distribution is
symmetric\footnote{Measurements of velocity increments in the solar
  wind  show asymmetric PDFs, which has lead
  \citet{2004GeoRL..3116807B} to add a cubic term to the expression in
  eq. (\ref{eq:tsallis}), in order to form a 
  ``generalized'' (non-symmetric) Tsallis distribution that can be
  used to describe skewed PDFs.}.
Varying the parameter $q$ in the Tsallis
distribution one obtain distributions that range from Gaussian
to ``peaky'' distributions with large tails. The parameter $q$ is
closely related to the kurtosis (fourth order one-point moment) of
the PDF, and similarly the parameter $w$ is related to the variance of
the PDF. However, they are not the same, while fitting a Tsallis
distribution to a PDF, the two parameters are calculated
simultaneously, therefore $w$ will depend on the value of $q$ and vice
versa, while the dispersion of a distribution is independent of the
kurtosis. In the next section we discuss how this difference might
play in our favor. 
Furthermore, the Tsallis distribution is often used in the context of
non-extensive statistical dynamics, it was originally derived
\citep{1988JSP....52..479T} from an entropy generalization to
extend the traditional Boltzmann-Gibbs statistics to multifractal
systems (such as the ISM). The Tsallis distribution reduces to the
classical Boltzmann-Gibbs (Gaussian) distribution in the limit of
$q\rightarrow 1$. 
It provides a physical foundation to other functions that have been
used to model empirically velocity PDFs in space plasmas \citep[such
as the kappa function, see][]{1997GeoRL..24.1151M,
  2002Ap&SS.282..573L}.
However, the physical interpretation of ISM simulations/observations 
PDFs in the light of such statistical dynamics is beyond the scope of
this paper.

Turbulence intermittency manifests itself in non-Gaussian
distributions, thus one can hope to establish a link between the
properties of turbulence and the Tsallis distributions parameters
obtained from simulations and/or observations. This provides an
alternative/complementary method to higher order statistics
\citep{2009ApJ...693..250B}, or to other PDF analysis
\citep[such as those in][where the kurtosis and/or skewness of the
distributions is measured]{1994ApJ...436..728F, 2000ApJ...535..869K},
all of which are sensitive to the non-Gaussian nature of real
turbulence.

Following the procedure presented by Burlaga and collaborators in what
follows we make
histograms of the distribution of fluctuations from our simulations. 
But, in contrast to their studies, where only the magnetic field
variations were studied at different times, we consider the spatial
variations of density, velocity and magnetic field intensity at a
given time, as a function of a spatial scale (or lag) $r$. To make the
fits and present the results we remove the mean value of the
increments and scale them to units of the standard deviation of each
PDF. In other words,  we have used
$\Delta f(r)=[f(r)-\langle f(r)\rangle_{\mathbf{x}}]/\sigma_{f}$, with
$f(\mathbf{x})=\rho(\mathbf{x}+r)-\rho(\mathbf{x})$,
$v_x(\mathbf{x}+r)-v_x(\mathbf{x})$, 
$v_y(\mathbf{x}+r)-v_y(\mathbf{x})$, 
$v_z(\mathbf{x}+r)-v_z(\mathbf{x})$, 
$B_x(\mathbf{x}+r)-B_x(\mathbf{x})$,
$B_y(\mathbf{x}+r)-B_y(\mathbf{x})$, 
and $B_z(\mathbf{x}+r)-B_z(\mathbf{x})$, where $\langle ...\rangle_{\mathbf{x}}$  stands for spatial average.
We have neglected the anisotropy introduced by the magnetic field
\citep[see][]{2009ApJ...701..236C}, the lag 
$r$ is considered along each of the three cardinal directions and the
resulting histograms include variations in all of them.

\section{MHD simulations}
\label{sec:tsallis-mhd}

We have taken a reduced subset of the MHD simulations of
 fully-developed (driven) turbulence of \citet{2007ApJ...658..423K}.
We use the data output of the simulations to study the resulting PDFs
of density, magnetic field and velocity in different turbulence
scenarios. 
 
The code used (described in detail in \citealt{2007ApJ...658..423K})
solves the ideal MHD equations in conservative form:
\begin{eqnarray}
\frac{\partial \rho}{\partial t}+\mathbf{\nabla \cdot}
\left(\rho\mathbf{v} \right) = 0, \label{eq:continuity}\\
\frac{\partial \rho\mathbf{v}}{\partial t}+\mathbf{\nabla
  \cdot}\left[\rho\mathbf{vv}+\left(p+\frac{B^2}{8\pi}\right)\mathbf{I}
  -\frac{1}{4\pi}\mathbf{BB} \right]=\mathbf{f},\label{eq:momentum}\\
\frac{\partial \mathbf{B}}{\partial t}-\mathbb{\nabla
  \times}\left(\mathbf{v \times B} \right)=\mathbf{0},\label{eq:induction}
\end{eqnarray}
along with the additional constraint $\mathbf{\nabla \cdot B}=0$ in a
Cartesian, periodic domain.
An isothermal equation of state $p=c_s^2 \rho$ is used, 
where $\rho$ is the mass density, $\mathbf{v}$ the velocity,
$\mathbf{B}$ the magnetic field, $p$ the gas pressure, and $c_s$ the
isothermal sound speed. The term $\mathbf{f}$ in the equation of
momentum (\ref{eq:momentum}) is a large scale driving. 
The driving is purely solenoidal, applied in Fourier space at a fixed
wave-number $k=2.5$ (a scale $2.5$ times smaller than the size of the 
computational box).
The rms velocity is maintained close to unity, so that $v$ is in units
of the rms velocity of the system, the Alfv\'en velocity
$v_A=\mathbf{B}/(4\pi\rho)^{1/2}$ is also in the same units. The
magnetic field is of the form
$\mathbf{B}=\mathbf{B}_{\mathrm{ext}}+\mathbf{b}$, that is  a uniform
background field $\mathbf{B}_{\mathrm{ext}}$ plus a fluctuating part
$\mathbf{b}$ (initially equal to zero). 
As it is customary, the units are normalized such that the Alf\'en
velocity $v_A=B_{\mathrm{ext}}/(4\pi\rho)^{1/2}=1$ and the
mean density $\rho_0=1$. This way, the controlling parameters are the
sound, and the Alfv\'en speeds (i.e. the gas pressure and the magnetic
pressure, respectively), yielding any combination of subsonic or
supersonic, with sub-Alfv\'enic or super-Alfv\'enic regimes of
turbulence. For this paper we explore the different combinations of
parameters summarized in Table \ref{tab:models}.
\begin{deluxetable}{cccccl}
\tablecaption{Parameters of the simulations\label{tab:models}}
\tablewidth{0pt}
\tablehead{
\colhead{Model} & \colhead{$p$}& \colhead{$B_{\mathrm{ext}}$}& 
\colhead{$\mathcal{M}_s$\tablenotemark{a}}& 
\colhead{$\mathcal{M}_A$\tablenotemark{b}}& \colhead{Description}}
\startdata
1 & 0.1 & 0.1 & 2.0 & 2.0 & Supersonic \& super-Alfv\'enic \\
2 & 0.1 & 1.0 & 2.0 & 0.7 & Supersonic \& sub-Alfv\'enic   \\
3 & 1.0 & 0.1 & 0.7 & 2.0 & Subsonic   \& super-Alfv\'enic \\
4 & 1.0 & 1.0 & 0.7 & 0.7 & Subsonic   \& sub-Alfv\'enic
\enddata
\tablenotetext{a}{Sonic Mach number, defined as
  $\mathcal{M}_s=\langle v/c_s\rangle$ (averaged over the entire
  computational domain).} 
\tablenotetext{b}{Alfv\'en Mach number, defined as
  $\mathcal{M}_A=\langle v/v_A\rangle$ (averaged over the entire
  computational domain).}
\end{deluxetable}

We should note that it has been recently found that a compressive
component of the driving can have an important effect on the
statistical properties of the flow \citep{2008PhRvL.101s4505S,
  2008ApJ...688L..79F, 2009A&A...494..127S}. This is an interesting
effect that require further study, but for the time being we will
restrict ourselves to the usual divergence-free driving.
 
\section{Statistics of three-dimensional data}
\label{sec:statistics}

\begin{figure*}
\centering
\includegraphics[width=0.75\textwidth]{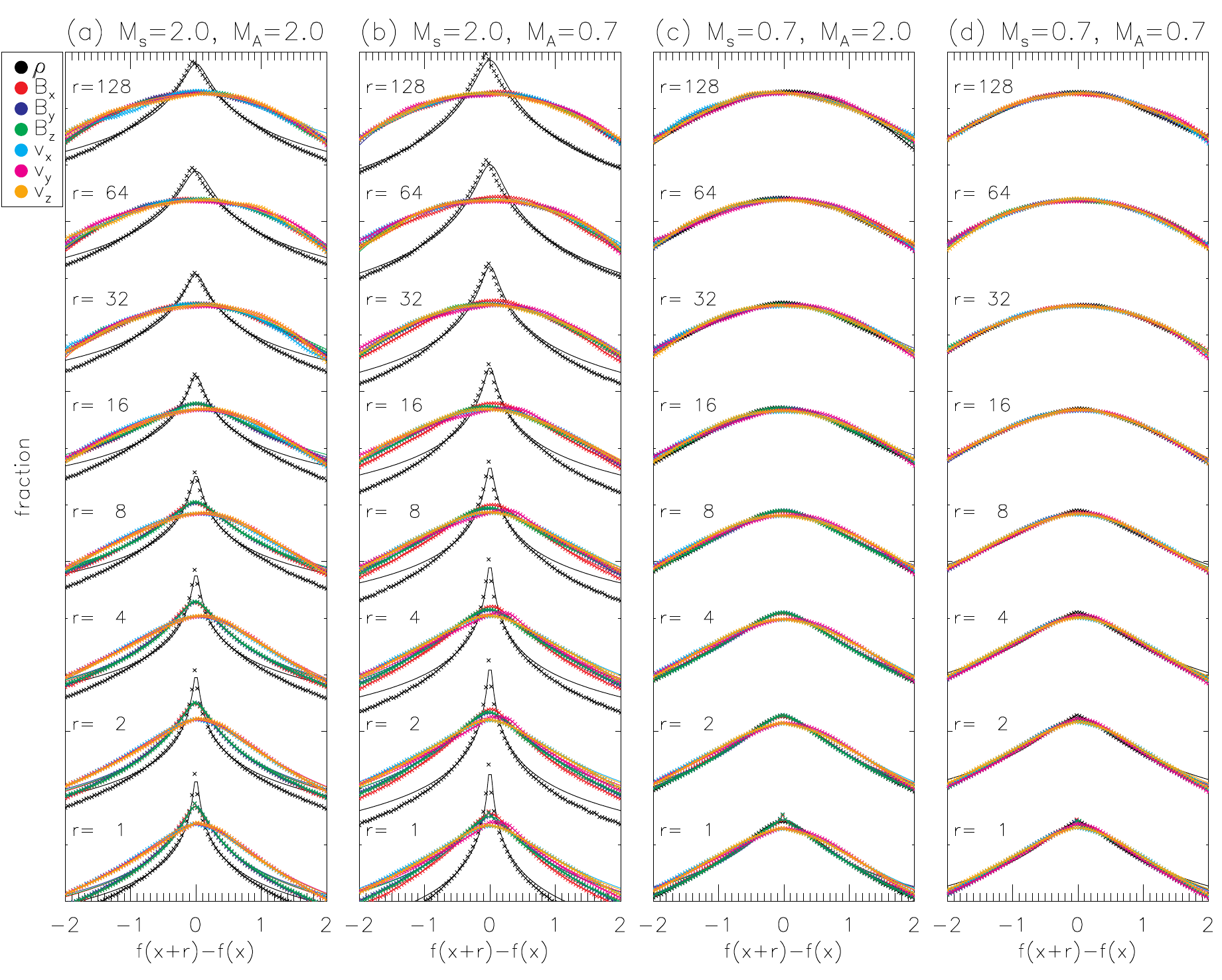}
\caption{PDFs of fluctuations for different spatial lags $r$, for all
  the models. The symbols represent data from the numerical
  simulations and the lines are the fits with a q-Gaussian (Tsallis)
  distribution. The different MHD variables are color coded according
  to the legend at the upper left corner, and the value of the lag is
  indicated at the left of each set of curves.}
\label{fig:pdf3d}
\end{figure*}

Below, we analyze the results of our simulations to test whether the
statistics of the PDFs of the fluctuations of the variables can be
described by the  Tsallis formalism. To to this we fit to each
histogram a Tsallis distribution with the Levenberg-Marquardt
algorithm \citep{1992nrfa.book.....P}, the histograms and 
the fits are presented in Figure \ref{fig:pdf3d}, the symbols are the
data from the simulations, and the lines are the corresponding
fits. For visual purposes we have shifted vertically the results by
successive factors of $100$ for each different lag.
Similarly to the results from solar wind observations
\citep{2004GeoRL..3116807B,2004JGRA..10912107B, 2005PhyA..356..375B,
  2005JGRA..11007110B,  2006ApJ...642..584B, 2006ApJ...644L..83B,
  2006PhyA..361..173B, 2007ApJ...668.1246B, 2007JGRA..11207206B,
  2009ApJ...691L..82B}, the magnetic field components, as
well as the density and velocity components, show kurtotic PDFs that
become smoother as we increase the scale-length (lag).
One difference is the fact that the velocity increments in
our simulations are symmetric, as opposed to the skewed
distributions observed in the solar wind at small scales.
It is also interesting that (only) for supersonic turbulence density
distributions are quite distinct from other quantities. Such
behavior resembles the power spectrum, where the magnetic field and
velocity for supersonic turbulence have steep spectra regardless of
sonic Mach number, while density becomes shallower as the Mach number
increases \citep[see for instance][]{2005ApJ...631..320E}. 
Such shallower slopes of the density power-spectrum have been also
reported when the driving changes from solenoidal to compressive 
(potential) for a fixed ($>1$) sonic Mach number
\citep{2009arXiv0905.1060F}.
This means that shocks in supersonic turbulence, which
create more small-scale structure in density, yield not only shallower
spectra, but also a larger degree of intermittency. At the same time,
such shocks can be formed easier if the turbulence driving mechanism
has compressive modes.

\begin{figure*}
\centering
\includegraphics[width=0.75\textwidth]{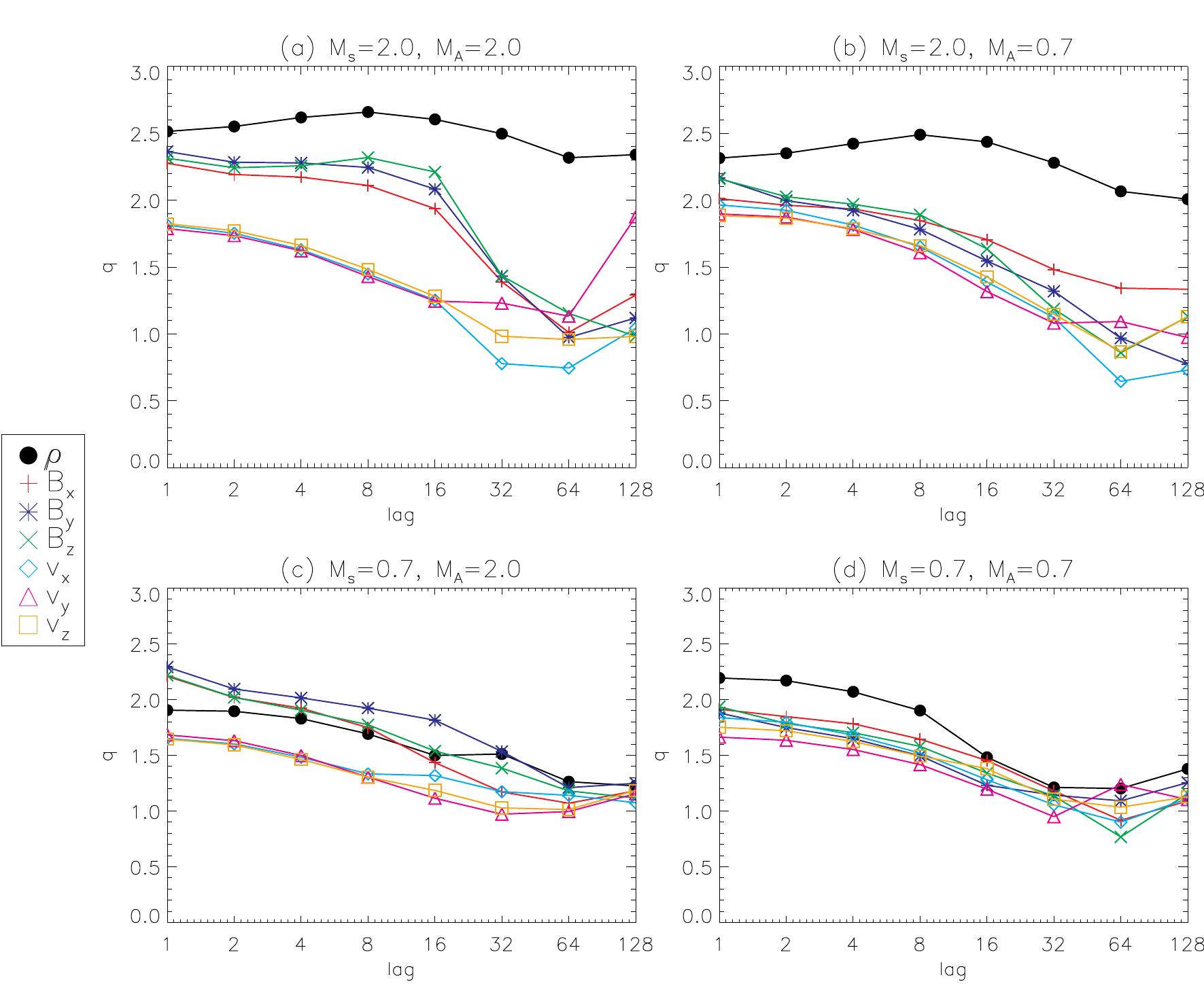}
\caption{Values of the $q$ parameter from the fits shown in Figure
  \ref{fig:pdf3d}, the symbols (and colors) indicate the MHD
variable used, according to the legend at the left of the plots.}
\label{fig:q3d}
\end{figure*}

\begin{figure*}
\centering
\includegraphics[width=0.75\textwidth]{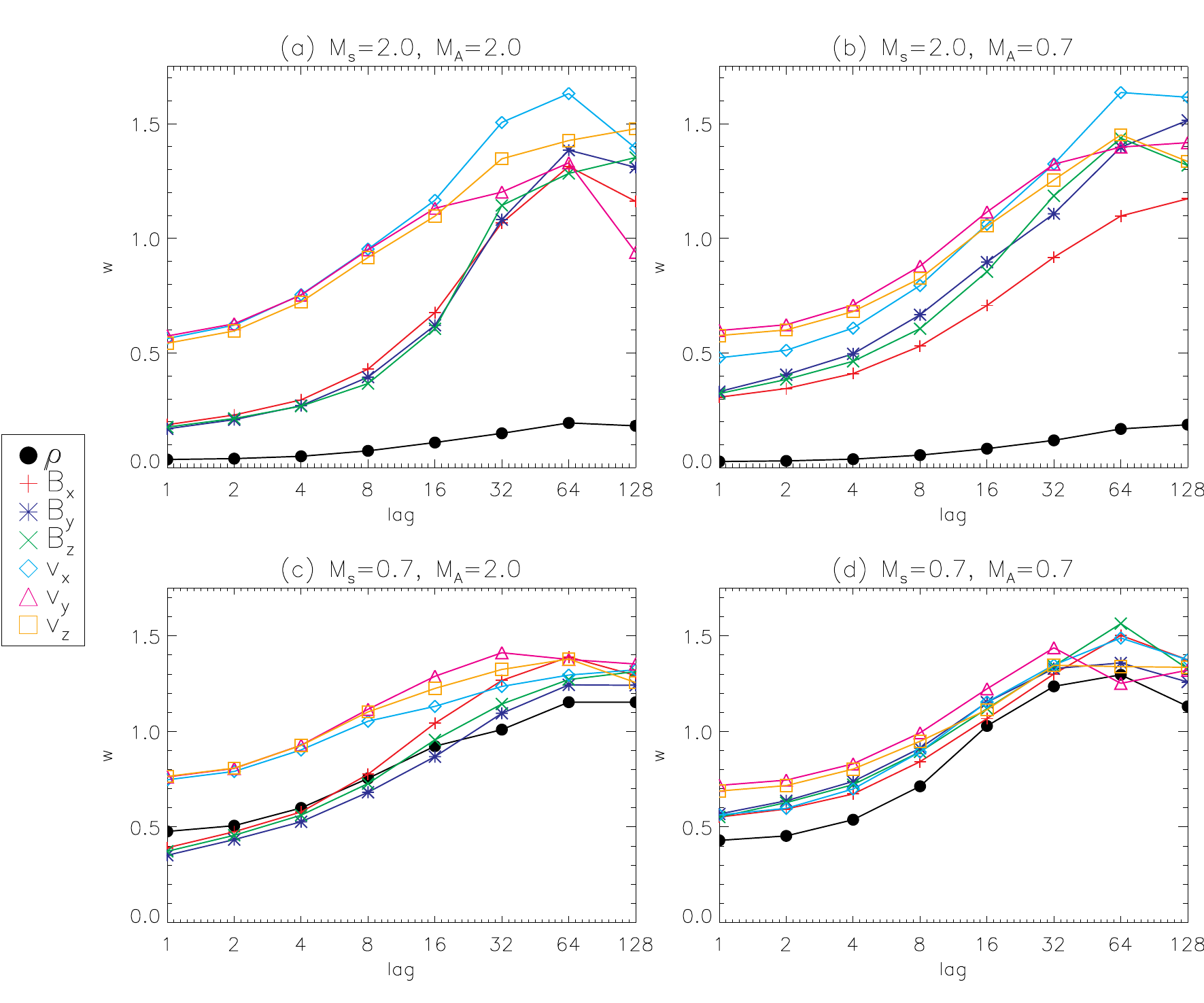}
\caption{Values of the $w$ parameter from the fits shown in Figure
  \ref{fig:pdf3d}, the symbols (and colors) indicate the MHD
variable used, according to the legend at the left of the plots.}
\label{fig:w3d}
\end{figure*}

We can observe from Figure \ref{fig:pdf3d} that the Tsallis
distributions fit  remarkably well the PDFs in the simulations. Only
small departures are evident at the tails of the PDFs, in
particular for density in models 1 and 2  (supersonic turbulence). 
The values of the $q$ and $w$ parameters from the fits are plotted as
a function of the separation $r$ in Figures \ref{fig:q3d}, and
\ref{fig:w3d}, respectively.

In Figures \ref{fig:q3d} and \ref{fig:w3d} we see the same general
trends found in solar wind studies
\citep{2004GeoRL..3116807B,2004JGRA..10912107B, 2005PhyA..356..375B, 
  2005JGRA..11007110B,  2006ApJ...642..584B, 2006ApJ...644L..83B,
  2006PhyA..361..173B,  2007ApJ...668.1246B, 2007JGRA..11207206B,
  2009ApJ...691L..82B}  and electron MHD simulations
\citep{2009ApJ...701..236C}: a $q$ parameter that decreases, along
with a $w$ value that increases as we take larger separations.
In other words, the PDFs become more kurtotic, and narrower, as we go
to small scales, while at large scales the PDFs are more Gaussian and
wider.
The same behavior (i.e. a decreasing kurtosis for an increasing lag) has
been found also for velocity centroids in simulations and observations
\citep[see ][]{1994ApJ...436..728F, 1998ApJ...504..889L,
  2000ApJ...535..869K, 2009arXiv0905.1060F}. 
In fact, we have computed the kurtosis and variance of the PDFs (not
shown here) and found a similar  monotonic decrease and increase,
respectively. Both the kurtosis and $q$ tend to their Gaussian
value ($3$ and $1$, respectively) as the separation tends to the
injection scale.

It is noticeable that the density PDFs in the supersonic regime have a
very distinct shape than those of the velocity  and magnetic field
components. In the models of supersonic turbulence they are much 
narrower, and more peaked than in subsonic turbulence. They also show a
shallower slope in the $q$ and $w$ vs $r$ plots.
We can also notice slightly larger values of $q$ (and lower $w$) for
the velocity components with respect to the magnetic field components,
more pronounced for high Alf\'en Mach numbers.

\section{Statistics of observable data: column density}
\label{sec:column}

The measures of non-Gaussianity presented in the previous section are
interesting in their own right, and might provide us with insights to
the intermittency of MHD turbulence from
simulations. However, observations of the ISM cannot provide direct
three-dimensional information of the density, velocity, and magnetic
fields. For instance, from spectroscopic observations we can access only the
velocity distribution of the emitting material at a given position (or
positions) in the plane of the sky.

\begin{figure*}
\centering
\includegraphics[width=0.75\textwidth]{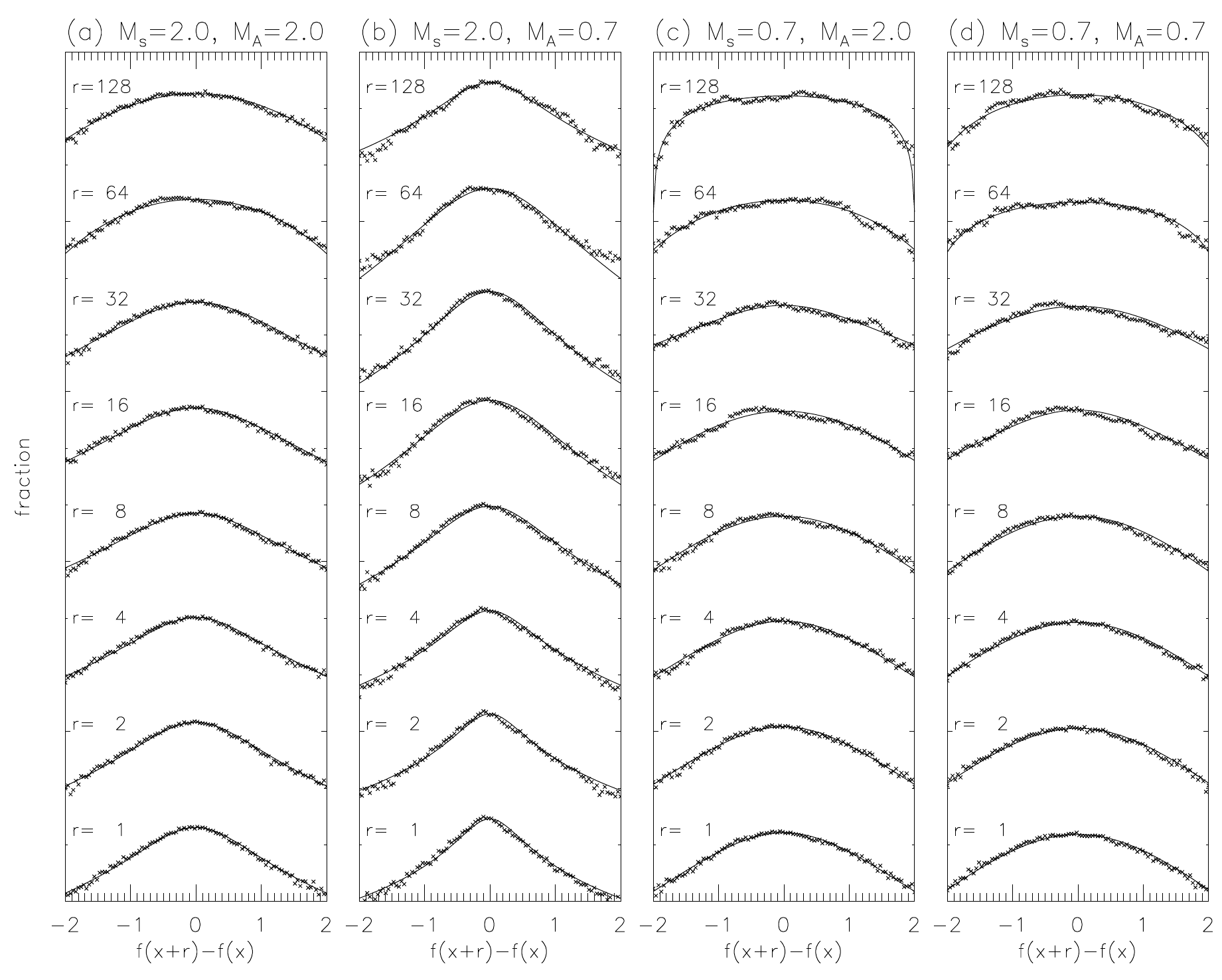}
\caption{PDFs of column density fluctuations for different spatial
  separations (``{\it x}'' symbols), and Tsallis fits ({\it solid}
  lines).} 
\label{fig:pdf2d}
\end{figure*}

With the simulations presented in the previous section we have
constructed (2D) maps of column density, assuming that the emitting
material is optically thin, and with an emissivity linearly
proportional to the density (e.g. \ion{H}{1}).
From the two-dimensional column density maps we constructed PDFs
of column density  increments, and following the same procedure we
described before, we made fits to Tsallis distributions. The PDFs and
the fits are presented in Figure \ref{fig:pdf2d}.

From the figure we can confirm the same general behavior of the PDFs,
which become more Gaussian as we take larger separations. We can see
that the PDFs are still consistent with Tsallis distributions,
although the dependence on the lag is more subtle than with the 3D
fields used in the previous section.

\begin{figure}
\centering
\includegraphics[width=0.85\columnwidth]{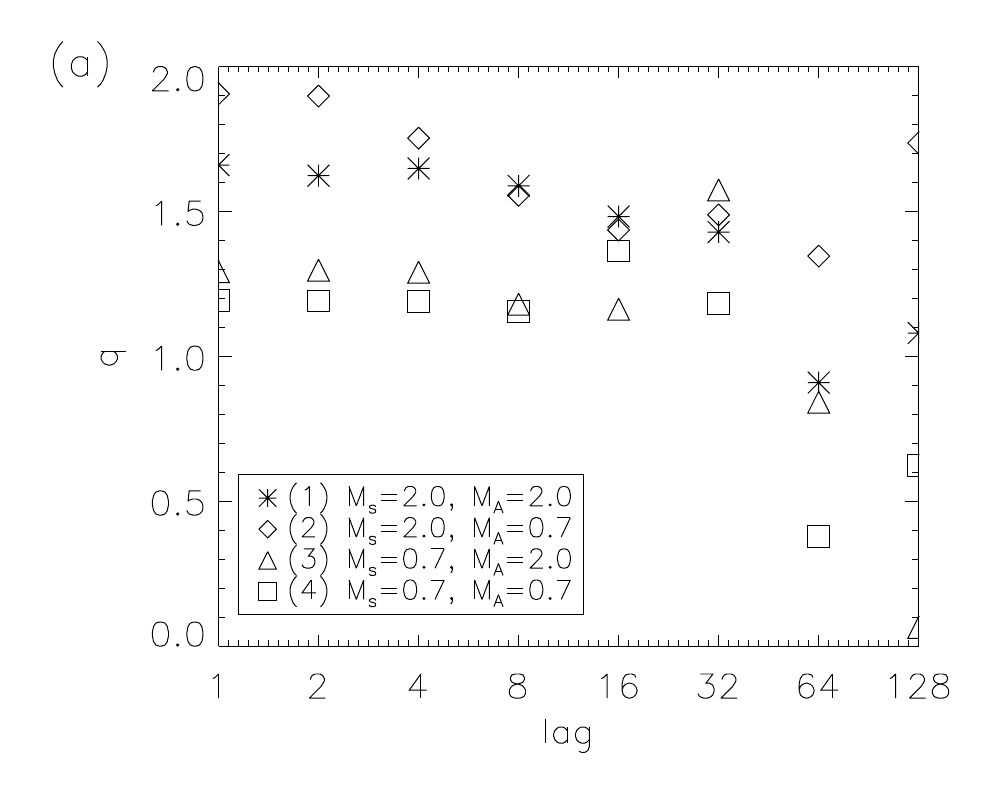}\\
\includegraphics[width=0.85\columnwidth]{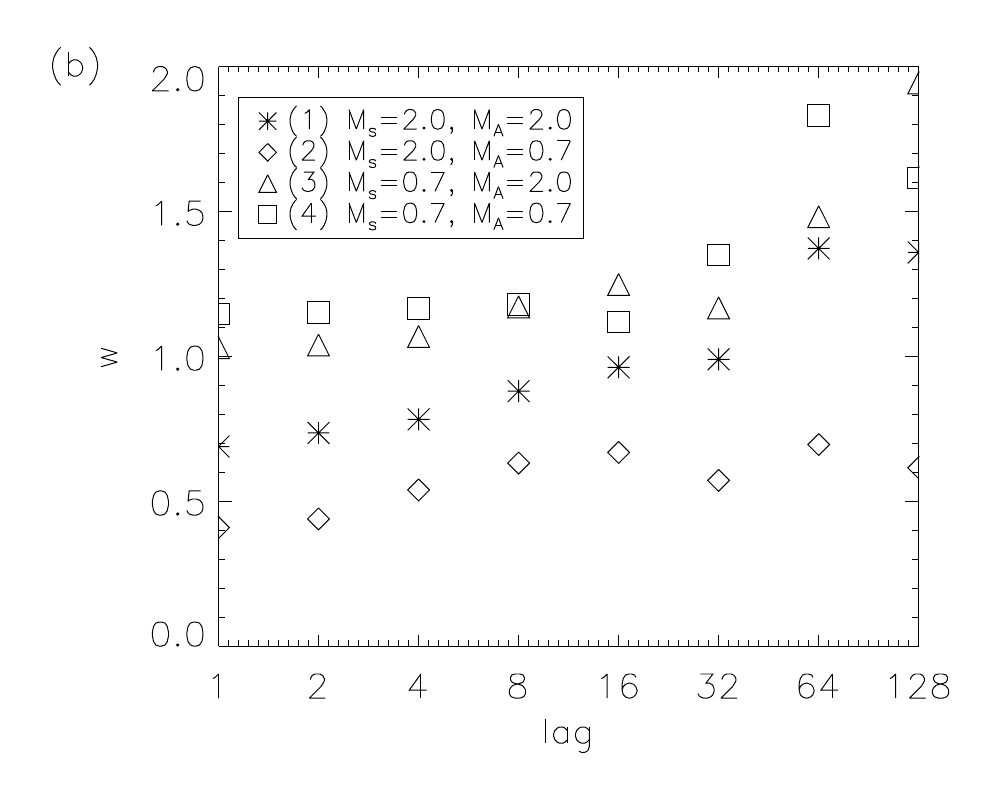}
\caption{Parameters q and w from d the Tsallis distribution fits shown in
  Figure \ref{fig:pdf2d}.}
\label{fig:q2d}
\end{figure}

In Figure \ref{fig:q2d} we present the parameters $q$ and
$w$ as obtained from the fitting procedure. Both panels of Figure
\ref{fig:q2d} confirm the decrease of $q$ along with an
increase of the width of the PDFs ($w$) with the lag.
From the figure it is also evident that the fitting
parameters can clearly distinguish between the supersonic and subsonic
models. However, this can be done with other measures, for instance
the variance of density or column density  \citep{1997ApJ...474..730P,
  2003A&A...398..845P, 2008ApJ...688L..79F, 2009arXiv0905.1060F}, or
the skewness and kurtosis of density and column density PDFs
\citep{2007ApJ...658..423K, 2009ApJ...693..250B}.
A more difficult task is to extract information of the Alfv\'{e}n
Mach number. 
One can see from Figure  \ref{fig:q2d} that the Tsallis parameters
depend strongly on the sonic Mach number, and less so on the Alfv\'en
Mach number. However, an encouraging clear dependence on  the Alfv\'en
Mach number is evident in the supersonic models, in particular for
$w$ (lower panel of Figure \ref{fig:q2d}), where 
all the $\mathcal{M}_A=2$ points lie above the $\mathcal{M}_A=0.7$
ones. But, for the subsonic models ($\mathcal{M}_s=0.7$) it not clear
at  all, and the behavior with the Alf\'en Mach number is the
opposite, higher $w$ values for sub-alfv\'enic models.

\section{Discussion}
\label{sec:discussion}

We found that the Tsallis expression [see Eq.~(\ref{eq:tsallis})]
describes well the statistics of the PDFs of our 3D-MHD numerical
simulations. This is welcome news for both the ISM, where we propose
to use Eq.~(\ref{eq:tsallis}) as a new tool to study turbulence, and
for the solar wind studies, where the Tsallis expression was used, but
its numerical testing was limited.

We have to add that the present work is intended only as a proof of
concept. While the Tsallis PDFs do provide physical grounds to
interpret the distributions obtained in MHD simulations and/or
observations, it is not clear whether the formalism of non-extensive
statistical dynamics is applicable to the turbulent ISM.
This, we feel
deserve a separate study. However, the fact that the expression
provides both the fit to numerical MHD simulations and experimental
data, and moreover seems to distinguish the Alfv\'en Mach number in
some cases, is encouraging.

In regard to observational quantities, we should note that PDFs of
velocity centroids (as opposed to column density, which is what we
ave used in the present paper) have been employed to characterize the 
intermittency both in numerical models \citep{1994ApJ...436..728F,
2000ApJ...535..869K, 2009arXiv0905.1060F} and molecular cloud
observations (for instance $\rho$-Ophiuchi in
\citealt{1998ApJ...504..889L}; Polaris and Taurus in
\citealt{2008A&A...481..367H}). There is always a trade off while
choosing different quantities to measure turbulence. On the one hand
velocity is a dynamical quantity that is easy to relate with
turbulence theories, however (at least in the case of power-spectra) is
trickier to obtain, and in fact (for power-spectra) velocity centroids fail
to trace the velocity statistics in strongly supersonic turbulence
\citep[see for instance][]{2005ApJ...631..320E}. On 
the other hand density fluctuations are not only produced from
turbulence, but are more robustly obtained from observations
\citep[e.g.][]{2008ApJ...686..350L}. We must stress that
the healthy approach is to use these tools, and others as well, as
complementary.


The most important finding of this paper is the dependence of the
$q$ and $w$ parameters entering Eq.~(\ref{eq:tsallis}) on both sonic
and Alfv\'{e}n Mach number of the turbulence. These numbers are
essential for many astrophysical processes from star formation
\citep[see][]{2007ARA&A..45..565M}, magnetic diffusion enabled by
reconnection \citep{2005AIPC..784...42L}, propagation of heat 
\citep{2001ApJ...562L.129N, 2006ApJ...645L..25L}, cosmic rays
\citep{2004ApJ...614..757Y} and dust grains acceleration 
\citep{2004ApJ...616..895Y}.

By now several ways of estimating the Alfv\'{e}n and sonic Mach numbers
have been proposed (see Kowal et al. 2007), all with their own
limitations and uncertainties. Thus the new approach here based on
Eq.~(\ref{eq:tsallis}) is a welcome addition to the existing tools.

We should add however, that a number of turbulence
parameters, which are as important as these Mach numbers, are
not being explored in this work. The scale of turbulence, for
instance, has important consequences for star formation
\citep[e.g.][]{2000ApJ...535..887K, 2004RvMP...76..125M} as well as for
dynamo theory \citep[e.g.][]{2005AN....326..400B, 2005PhR...417....1B,
  2009A&A...498..335H}

\section{Summary}
\label{sec:summary}
We have used three-dimensional MHD simulations to study the
distribution functions of increments of density,  velocity, and
magnetic field. In addition, we studied synthetic column density
maps, which can be directly compared with observations. 
 The results can be summarized as follows. 
\begin{enumerate}
  \item The PDFs of each of all the quantities is well described by a
    Tsallis distribution. Such distributions are symmetric, and
    characterized basically by three parameters: an amplitude $A$, a
    width $w$, and the entropic index $q$ that relates to the shape
    (large values yield peaked distributions, while with small values
    they approximate to a Gaussian). 
  \item The general trend of the PDFs for different increments are
    consistent with previous studies, of solar wind observations. That
    is, the PDFs are close to a Gaussian distribution for large
    separations, and become more kurtotic and narrower (i.e. a larger
    degree of intermittency) for small separations.
\item While the PDFs of (3D) density are quite similar to those of
  magnetic field and velocity in subsonic turbulence (regardless of
  Alfv\'en Mach number), for supersonic turbulence (also independently
  of $\mathcal{M}_A$) they are noticeable narrower and more
  peaked. That is,  the fits of the PDFs to a Tsallis distribution
  have smaller $w$, and larger $q$ parameters. 
\item The column density maps show smoother distributions
  compared to those of density in 3D, however, we found that the $q$
  values are larger for supersonic turbulence (and lower $w$ as well),
  regardless of the Alfv\'en Mach number. 
\item The $w$ parameter of the Tsallis fits to the PDFs of column
  density increments seems to show a sensitivity to the Alfv\'en Mach
  number, at least in the cases of supersonic turbulence.
  With the limited data set used here is hard to conclude if this
  behavior is general, and whether it is useful to study the
  magnetization of ISM turbulence. Clearly, future studies of the PDFs
  in ISM turbulence (with a wider parameter coverage) are necessary.
\end{enumerate}

\acknowledgments We thank the anonymous referee for his/her many
useful suggestions, also G. Kowal for providing the data cubes of the
MHD simulations analyzed in this paper.
AE acknowledges support from the DGAPA (UNAM) grant 
IN108207, from the CONACyT grants 46828-F and 61547, and from the
``Macroproyecto de Tecnolog\'\i as para la Universidad de la
Informaci\'on y la Computaci\'on'' (Secretar\'\i a de Desarrollo
Institucional de la UNAM). AL acknowledges NSF grants AST 0808118 and
the support of the NSF Center for Magnetic Self-Organization. 

\bibliography{Master}

\end{document}